\documentclass[twoside,11pt,a4paper,final,twocolumn]{revtex4}

\usepackage[dvips]{graphicx}
\usepackage{epsfig}
\usepackage{amsfonts}
\usepackage{colordvi}

\begin{document}

\title{Analysis of complete positivity conditions for quantum qutrit channels}

\author{Agata Ch\c{e}ci\'nska} \email{agata.checinska@fuw.edu.pl}
\affiliation{Instytut Fizyki Teoretycznej, Uniwersytet Warszawski,
  Warszawa 00--681, Poland}
\affiliation{ICFO-Institut de Ciencies Fotoniques, Mediterranean Technology Park,
08860 Castelldefels (Barcelona), Spain}

\author{Krzysztof W\'odkiewicz}
 \affiliation{Instytut Fizyki Teoretycznej,
  Uniwersytet Warszawski, Warszawa 00--681, Poland}

\affiliation{Department of Physics and Astronomy, University of New
  Mexico, Albuquerque, NM~87131-1156, USA}
\date{\today}

\begin{abstract}

We present an analysis of complete positivity (CP) constraints on
qutrit quantum channels that have a form of affine transformations
of generalized Bloch vector. For diagonal (damping) channels we
derive conditions analogous to the ones that in qubit case produce
tetrahedron structure in the channel parameter space.

\end{abstract}
\maketitle
\section{Introduction}

The analysis of quantum channels - completely positive trace
preserving maps - is one of the crucial points for quantum
information theory.
 Quantum channels correspond to processes that physically may take place and lead to evolution of a quantum state \cite{zyczkowski,chuang}.
  For qubit case, we already know the analysis of quantum qubit channels \cite{ruskai}. This analysis gives us a connection between
  the physical evolution of the system written via Bloch equations and quantum channel
  formalism \cite{ruskai}. Hence one can derive mathematical conditions on parameters that appear in Bloch formalism.
  These mathematical conditions are the consequence of the fact that each physical process is a completely positive map.

In our work, we aim at presenting similar analysis for qutrit
channels. Qutrit states are states belonging to three dimensional
Hilbert space and, in analogy to qubit case, one can use
generalized Bloch formalism to describe their evolution
\cite{byrd}. Generalized Bloch equations, that describe for
instance three level atoms, may present the physical context for
the evolution of qutrit quantum state. In analogy to qubit case,
we investigate the evolution of generalized Bloch vector that
evolves within a Bloch ball. As a natural choice, we investigate
qutrit channels of the general form of linear transformation on
the qutrit Bloch vector - studying affine transformations on
qutrit Bloch vectors. We parameterize qutrit channels and then
derive conditions for channel parameters in order to obtain
physical transformations: completely positive maps (CPM). In qubit channel analysis
 a tetrahedron structure of completely positive maps appears (for qubit unital channels), 
we show analogous analysis 
for qutrit channels, for which more sophisticated channel geometry emerge. This new result 
on qutrit channels can be linked with the analysis of bipartite qutrit states 
via Jamiolkowski isomorphism.

\section{State description}

Let us first recall the idea behind the Bloch formalism. This, in
qubit case, corresponds to the choice of representation of the
qubit state density operator: the basis of Pauli matrices
$\sigma_i$ \cite{chuang,daffer}
\begin{equation}
\rho_{qb}=\frac{1}{2}(\mathbb{I}+\vec{b}\,\vec{\sigma}),
\end{equation}
where $\vec{b}$ is a three dimensional, real Bloch vector, describing the qubit state and satisfying
$\vec{b}^2\leq1$ (equality for pure states). Qubit states occupy entirely the Bloch ball. In a similar way we
can represent a qutrit state, a state belonging to three dimensional Hilbert space. In qutrit case the choice of
representation is set to be the basis of Gell-Mann matrices $\lambda_i$, the generators of $SU(3)$ group \cite{klimov}
\begin{equation}
\rho_{qt}=\frac{1}{3}(\mathbb{I}+\sqrt{3}\,\vec{n}\,\vec{\lambda}).
\end{equation}
Here $\vec{n}$ is a generalized Bloch vector, real and eight dimensional. Qutrit states can
be characterized by condition $\vec{n}\leq1$. However, qutrit case is more sophisticated: pure states are states
for which two conditions are satisfied \cite{klimov,byrd}:
\begin{equation}
\vec{n}^2=1,\qquad\vec{n}*\vec{n}=\vec{n},\label{purestates}
\end{equation}
where $*$-product is defined as
$(\vec{A}*\vec{B})_i=d_{ijk}A_jB_k$, with $d_{ijk}$ being totally
symmetric tensor \cite{byrd}. Qutrit states belong to a
generalized Bloch ball, qutrit pure states belong to the unit
sphere $\mathcal{S}^7=\{\vec{n}\in\mathrm{R}^8\,:\,\vec{n}^2=
1\}$. However, physical qutrit states do not occupy entirely the
generalized Bloch ball. The pure qutrit states (states satisfying
conditions (\ref{purestates})) form a subset of the unit sphere.
They can be parametrized with 4 parameters \cite{klimov} as
follows
\begin{eqnarray}
|\Psi\rangle\,&=&\,e^{\imath\chi_1}\sin{\theta}\cos{\phi}|0\rangle\,+\,e^{\imath\chi_2}\sin{\theta}\sin{\phi}|1\rangle\,+\nonumber\\
&\  &+\,\cos{\theta}|2\rangle,
\end{eqnarray}
where $0\leq\theta,\phi<\frac{\pi}{2},\,0\leq \chi_1,\chi_2 < 2\pi$ and overall phase was omitted.
Hence, the set of pure qutrit states is a 4 dimensional subset of 7 dimensional sphere.

\section{Completely Positive Trace Preserving Maps}
  It was shown in \cite{choi} that physical transformations must not only be
  positivity preserving but there exist more subtle conditions to satisfy, these are called complete
  positivity (CP) conditions \cite{choi,zyczkowski}. The classification of qubit channels according to complete positivity is well known \cite{ruskai}.
  We want to present similar analysis for qutrit channels. \\
We represent physical system with Hilbert space $\mathcal{H}$. $\mathcal{B}(\mathcal{H})$ is the algebra of all
bounded operators on $\mathcal{H}$, a linear map
$\Phi\,:\,\mathcal{B}(\mathcal{H})\mapsto\,\mathcal{B}(\mathcal{H})$ is completely positive if for every
positive integer \textit{m} the map:
\begin{equation}
\Phi^{(m)}=\Phi\otimes\mathbb{I}^{(m)}:\mathcal{B}(\mathcal{H})\otimes\mathcal{M}^{(m)}\mapsto \mathcal{B}(\mathcal{H})\otimes\mathcal{M}^{(m)},\label{cp_def}
\end{equation}
is positive (where $\mathbb{I}^{(m)}$ is the identity operator on the algebra $\mathcal{M}^{(m)}$ of $m\times m$
complex matrices) \cite{choi}. Clearly, this amounts to saying that $\Phi$ acts on a subsystem A of a larger Hilbert space and there is a reservoir (or subsystem B) on which we act with unit operator $\mathbb{I}^{(m)}$. Here, we do not know the dimension of the reservoir and therefore $\Phi^{(m)}$ must be positive for any m.
It was shown that every CPM has an operator-sum or Kraus representation \cite{kraus}
\begin{equation}
 \Phi^{CPM}(\rho)=\sum_i \mathcal{K}_i\rho\mathcal{K}_i^{\dagger},
\end{equation}
with $\mathcal{K}_i$ being a set of Kraus operators satisfying $\sum_i\mathcal{K}_i^{\dagger}\mathcal{K}_i=\mathbb{I}$.\\
To evaluate whether a given transformation $\Phi$ (a linear map)
is completely positive we need to construct the so called
dynamical (or Choi) matrix of the size $N^2\times N^2$ (N is the
dimension of the system of interest). We will denote the dynamical
matrix with $D_{\Phi}$ \cite{choi,zyczkowski}. Dynamical matrix
represents uniquely channel action. We denote with $E_{jk}$
$N\times N$ matrix with 1 at position (j,k) and zeros elsewhere.
The map $\Phi$ is CPM iff
\begin{equation}
D_{\Phi}\,\equiv\,\sum_{i,\,j=1}^N\Phi(E_{ij})\otimes E_{ij},
\end{equation}
is positive semi-definite ($D_{\Phi}\,\geq\,0$).\\
Channel $\Phi$ must preserve hermiticity of density matrix and
therefore its dynamical matrix must be hermitian:
$D_{\Phi}\,=\,D_{\Phi}^{\dagger}$. Trace preserving of the density
operators means that the partial trace of $D_{\Phi}$ with respect
to the first subsystem (A) gives the unit operator for the second
subsystem: $Tr_A D_{\Phi}\,=\,\mathbb{I}$. To evaluate the entries
of dynamical matrix, we need to compute the action of the channel
$\Phi$ on $E_{jk}$. Once again, when its action is rewritten as
\begin{equation}
\Phi(\rho)_{\mu\nu}=\sum_{\sigma,\tau=1}^N \Phi_{\mu\nu,\sigma\tau}\rho_{\sigma\tau},
\end{equation}
 where coefficients $\Phi_{\mu\nu,\sigma\tau}$ characterize channel action, we see that we work with $N^4$ numbers.

 \subsection*{Channels versus states -  Jamiolkowski isomorphism}

 On the other hand, $N^2\times N^2$, positive and hermitian dynamical matrix $D_{\Phi}$ must correspond to a
 density operator acting on an $N^2$-dimensional Hilbert space. This correspondence is up to the normalization factor,
 since $Tr D_{\Phi}\,=\,N$. Hence $\rho_{\Phi}\,=\,\frac{1}{N}\,D_{\Phi}$ is a proper density matrix that we can write as
 \begin{eqnarray}
 \rho_{\Phi}&=&\frac{1}{N}D_{\Phi}=\frac{1}{N}\sum_{i,\,j=1}^N\Phi(|i\rangle\langle j|)\otimes|i\rangle\langle j|\nonumber\\
 &=&\frac{1}{N}D_{\Phi}=\frac{1}{N}\sum_{i,\,j=1}^N\Phi(E_{ij})\otimes E_{ij}.
 \end{eqnarray}
 The set of density operators defined by the dynamical matrices is only a subset of density matrices in $N^2$-dimensional
 Hilbert space, since dynamical matrices must satisfy $Tr_A D_{\Phi}\,=\,\mathbb{I}$. The fact that completely positive maps $\Phi^{CPM}$,
 which are represented uniquely by they dynamical matrices, correspond to states is known as the Jamiolkowski isomorphism \cite{jamiolkowski}. \\
Therefore, when analyzing quantum qutrit channels we can
reinterpret it as an analysis of two qutrit quantum states, in
qutrit case $N=3$, therefore $\rho_{\Phi}$ is a $9\times 9$
matrix.

\section{Quantum qubit channels}
\subsection{Bloch equations}
 To recall the qubit case analysis, we can start with a two-level quantum system and its evolution.
 The latter can be written by means of Bloch equations that are equations for components of Bloch
 vector $\vec{b}\,=\,(u,v,w)$. If we take, for instance, the decoherence of a two-level atom, 
these equations read
\begin{eqnarray}
\dot{u}&=&-\frac{1}{T_u}u\,-\,\Delta\,v,\nonumber\\
\dot{v}&=&-\frac{1}{T_v}v\,+\,\Delta\,u\,+\,\Omega\,w,\\
\dot{w}&=&-\frac{1}{T_w}(w-w_{eq})-\Omega\,w,\nonumber
\end{eqnarray}
and represent the evolution of the system. Here $\Omega,\,\Delta$
are Rabi frequency and detuning respectively. $\frac{1}{T_i}$
stand for decay rates for the atomic dipole ($i\,=\,u,v$) and
decay rate of the atomic inversion ($i\,=\,w$). These equations,
when put together, give rise to an affine transformation of the
qubit Bloch vector that is governed by the parameters listed
above. Any physical process amounts to a transformation of the
qubit state that is already a completely positive map. However,
\emph{not} every affine transformation of the (qubit) Bloch vector
will be a completely positive map.

\subsection{Affine transformations on qubit Bloch vectors}

The analysis of completely positive trace preserving maps on
$\mathcal{M}_2$ (complex two dimensional matrices) has been
studied extensively \cite{ruskai,daffer} and gives the answer to
the problem. Without loss of generality one can analyze qubit
channels that transform qubit Bloch vector according to
\begin{equation}
 \Phi^{qb}\,:\,\vec{b}\mapsto\vec{b}'=\Lambda^{qb}\vec{b}+\vec{t}^{qb},
\end{equation}
where matrix $\Lambda^{qb}=diag\{\Lambda^{qb}_1,\Lambda^{qb}_2,\Lambda^{qb}_3\}$
consists of damping eigenvalues $\Lambda^{qb}_i$ and $\vec{t}^{qb}=(t_1^{qb},t_2^{qb},t_3^{qb})$
is a translation. The image of the set of pure states ($\vec{b}^2\,=\,1$, Bloch sphere)
under such transformation is the ellipsoid
\begin{equation}
 (\frac{u'-t^{qb}_1}{\Lambda^{qb}_1})^2+(\frac{v'-t^{qb}_2}{\Lambda^{qb}_2})^2+
 (\frac{w'-t^{qb}_3}{\Lambda^{qb}_3})^2\,=\,1,
\end{equation}
with its center defined by $\vec{t}$ and its axes by $\Lambda_i$.
The set of conditions on both $\Lambda^{qb}_i$ and $t^{qb}_i$ can
be found in \cite{ruskai,daffer}. When we limit ourselves just to
diagonal qubit channels (meaning $\vec{t}^{qb}\,=\,0$), then the
set of allowed $\Lambda^{qb}_i$ forms a tetrahedron structure
\cite{ruskai,zyczkowski}. This structure reappears also in the
space of two qubit states.

\section{Quantum qutrit channels}
\subsection{Qutrit Bloch equations}

As before, we can start the analysis of transformations on qutrit
quantum states with the analysis of a three level atom for wich we
can write down Bloch equations. Three level atom is not the only
possible physical realization \cite{vallone,lanyon} but it is very
illustrative. The analog of the Bloch vector for the case of a
three level atom was in the beginning introduced as a (eight
dimensional, real) coherent vector $\vec{S}$ \cite{eberly}, which
components (denoted as $u,v,w$) were defined as
\begin{eqnarray}
u_{jk}&=&\rho_{jk}\,+\,\rho_{kj},\nonumber\\
v_{jk}&=&\imath(\rho_{jk}\,-\,\rho_{kj}),\\
w_{jk}&=&-\sqrt{\frac{2}{l(l+1)}}\nonumber\\
&\  &\times(\rho_{11}+\rho_{22}+...+\rho_{ll}-l\,\rho_{l+1,l+1}),\nonumber
\end{eqnarray}
with $1\leq j<k\leq 3$ and $1\leq l \leq 2$. Now, as an example of
the physical system we can take a  three level atom for which
nonzero dipole moments are between levels 1 and 2, and 2 and 3.
The atom interacts with the electric field (two electromagnetic
waves incident on the atom) and we assume that detunings are the
same ($\Delta_{12}=-\Delta_{23}=\Delta$). The corresponding Bloch
equations for coherent vector $\vec{S}$ are
\begin{eqnarray}
\dot{u}_{12}&=&\Delta v_{12}+\beta v_{13},\nonumber\\
\dot{u}_{23}&=&-\Delta v_{23}-\alpha v_{13},\nonumber\\
\dot{u}_{13}&=&\beta v_{12}-\alpha v_{23},\nonumber\\
\dot{v}_{12}&=&-\Delta u_{12}-\beta u_{13}+2\alpha w_1,\nonumber\\
\dot{v}_{23}&=&\Delta u_{23}+\alpha u_{13}-\beta w_1+\sqrt{3}\beta w_2,\nonumber\\
\dot{v}_{13}&=&-\beta u_{12}+\alpha u_{23},\nonumber\\
\dot{w}_{1}&=&-2\alpha v_{12}+\beta v_{23},\nonumber\\
\dot{w}_{2}&=&-\sqrt{3}\beta v_{23},
\end{eqnarray}
where $\alpha, \beta$ are related to two Rabi frequencies
\cite{eberly}. These equations are a generalization of the
equations we have seen in the qubit case.\\
In this work we use slightly different notation for the qutrit
vector - we already have introduced qutrit Bloch vector $\vec{n}$
related to the choice of Gell-Mann matrices basis (in some works
generalized Bloch vectors are also called coherent vectors
\cite{byrd}). These two vectors ($\vec{S}$ and $\vec{n}$) are of course equivalent.\\
Parameters that appear in the qutrit Bloch equations have physical
background, therefore the resulting affine transformation is a
completely positive map. However, we can as well ask the opposite
question: given an arbitrary affine transformation on qutrit Bloch
vector what are the conditions on its parameters which guarantee
complete positivity?

\subsection{Affine transformations of qutrit Bloch vectors}

Having in mind the question stated above, we will look at
transformations of qutrit Bloch vector that have a form
\begin{equation}
 \Phi\,:\,\vec{n}\mapsto \vec{n}'=\Lambda\vec{n}+\vec{t},\label{diag}
\end{equation}
where $\Lambda=diag\{\Lambda_1,...,\Lambda_8\}$ consists of 8
damping coefficients and $\vec{t}$ is an eight dimensional
translation. The image of the set of pure states under this
transformation is
\begin{equation}
 \sum_{i=1}^8 (\frac{n'_i-t_i}{\Lambda_i})^2\,=\,1,
\end{equation}
  together with the condition for $*$-product $\vec{n}*\vec{n}\,=\,\vec{n}$
\begin{equation}
 \frac{n'_i-t_i}{\Lambda_i}\,=\,d_{ijk}\frac{n'_j-t_j}{\Lambda_j}\frac{n'_k-t_k}{\Lambda_k}.
\end{equation}
On the other hand, parameters $\Lambda_i,\,t_i$ must satisfy
\begin{equation}
 \sum_i(\Lambda_i n_i\,+\,t_i)^2\,\leq\,1,
\end{equation}
according to the requirement $\vec{n}'^2\,\leq\,1$. However, complete positivity is a much stronger condition
than condition saying that we cannot exceed value 1 for the length of Bloch vector. The latter, in qubit case,
amounts only to statement that the density operator must be a positive definite operator. In qutrit case
however, it is even less than that - since not every point within the $\mathcal{S}^7$ sphere corresponds to
density operator. \\
Transformation (\ref{diag}) can be rewritten to give channel coefficients
$\Phi_{\mu\nu,\sigma\tau}$. To construct dynamical matrix $D_{\Phi}$, we apply the channel action to
$E_{jk}\,\mapsto\,\Phi(E_{jk})$, representing it in the basis of Gell-Mann matrices:
$E_{jk}=\frac{1}{\sqrt{3}}n^{jk}_{\alpha}\lambda_{\alpha}$ (where $\alpha\in\{0,...,8\}$,
$\lambda_0=\sqrt{\frac{2}{3}}\mathbb{I}$ and $n^{jk}_{\alpha}$ can be
interpreted as an analog of Bloch vector).\\
We will first look at the channels that consist only of damping matrix (diagonal channels) and do not have a translation. These channels are in fact unital, since they leave maximally mixed state unchanged (they are called bistochastic maps \cite{zyczkowski}). Later on, we will look at channels that include also translations of Bloch vector.

\subsection{CPM conditions for diagonal channels}

In qubit case, the action of the diagonal channel can be written as
\begin{equation}
\vec{b}\rightarrow\vec{b}'= \Lambda^{qb}\vec{b},\qquad
\Lambda^{qb}=diag\{\Lambda_1^{qb},\Lambda_2^{qb},\Lambda_3^{qb}\},
\end{equation}
 whereas for qutrits we have
\begin{equation}
 \vec{n}\rightarrow\vec{n}'=\Lambda\vec{n},\qquad
 \Lambda=diag\{\Lambda_1,...,\Lambda_8\}.
\end{equation}
In both cases, we assume that the nature of $\Lambda_i$ parameters is quasi-damping, hence $|\Lambda_i|\leq 1$.
This comes from the fact that $\vec{n}^2\,\leq\,1$ at all times, therefore, the change of any initial Bloch
vector will lead to a vector within the (generalized) Bloch ball. Dynamical matrix $D_{\Phi}$ for a qutrit channel of the
form (\ref{diag}) must be positive semi-definite in order to correspond to CPM. There are nine eigenvalues $d_i$ that
must be nonnegative to satisfy positivity of $D_{\Phi}$. The first six eigenvalues give rise to conditions that can be
written as {\small
\begin{equation}
 \begin{array}{ccc}
 1-\Lambda_8+\frac{3}{2}(\Lambda_4-\Lambda_5)&\geq &0,\\
 1-\Lambda_8-\frac{3}{2}(\Lambda_4-\Lambda_5)&\geq&0,\\
  1-\Lambda_8+\frac{3}{2}(\Lambda_6-\Lambda_7)&\geq&0,\\
 1-\Lambda_8-\frac{3}{2}(\Lambda_6-\Lambda_7)&\geq&0,\\
  1-\Lambda_8+\frac{3}{2}(\Lambda_1-\Lambda_2)+\frac{3}{2}(\Lambda_8-\Lambda_3)&\geq&0,\\
 1-\Lambda_8-\frac{3}{2}(\Lambda_1-\Lambda_2)+\frac{3}{2}(\Lambda_8-\Lambda_3)&\geq&0.
\end{array}\label{cond1}
\end{equation}}
These conditions alone lead to the set of allowed $\Lambda_i$ that
has a polyhedron like structure. In qubit case we have similar set
of equations for $\Lambda^{qb}$ that define the tetrahedron
structure. However, in qutrit case there are three remaining
inequalities (given by eigenvalues $d_7,\,d_8,\,d_9\,\geq\,0$)
which reveal coupling between all the parameters.
\begin{eqnarray}
d_7(\Lambda_1,...,\Lambda_8)&\geq&0,\nonumber\\
d_8(\Lambda_1,...,\Lambda_8)&\geq&0,\nonumber\\
d_9(\Lambda_1,...,\Lambda_8)&\geq&0.\label{cond2}
\end{eqnarray}
Because of their numerical complexity they are discussed in
Appendix. Matrix $D_{\Phi}$ is hermitian, therefore eigenvalues
$d_{7,8,9}$ must be real. For some cases, three conditions
(\ref{cond2}) reduce to just two (see Appendix).
 All the inequalities characterize the
set of allowed $\{\Lambda^{CPM}\}$, in other words, channel
parameters $\Lambda_i$ for which $\Phi$ is a CPM. The boundaries
of the set $\{\Lambda^{CPM}\}$ can be computed by analyzing values
of $\Lambda_i$ satisfying equations instead of inequalities given
by (\ref{cond1}) and (\ref{cond2}). In principle, parameters
$\Lambda_i$ can be time
dependant, still, conditions (\ref{cond1}) and (\ref{cond2}) must be satisfied for any time \emph{t} to have a CPM.\\
If we assume, for example, that
\begin{equation}
\frac{dn_i(t)}{dt}=\gamma_i\,n_i(t),
\end{equation}
then time evolution of the Bloch vector $\vec{n}(t)$ is given by:
\begin{equation}
n_i(t)=e^{\gamma_it}n_i(0).
\end{equation}
We can then identify $\Lambda_i\,=\,e^{\gamma_it}$ and conditions
on $\Lambda_i$ will impose conditions on $\gamma_i$. For this type
of evolution, one can write the Lindblad equation for qutrit
density operator $\rho(t)$, corresponding to the channel action.
Some more details on relation between complete positivity and
master equation and Lindblad operators one can find in
\cite{gorini,daffer,hall}.

\subsubsection*{The structure of the set $\{\Lambda^{CPM}\}$}

\begin{figure}
\begin{center}
\includegraphics[scale=0.4]{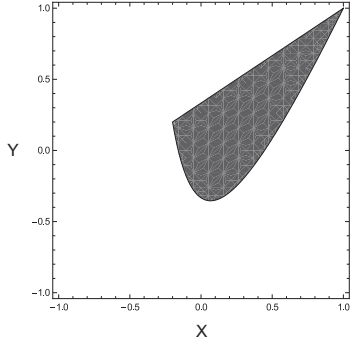}
\caption{The dark region in the figure shows these values of
parameters $\Lambda_1=\Lambda_2=Y,\;\Lambda_{i\neq 1,2}=X$ that
satisfy CP conditions. The qutrit channel has only the diagonal
(damping) part.} \label{one}
\end{center}
\end{figure}

\begin{figure}[h!]
\includegraphics[scale=0.4]{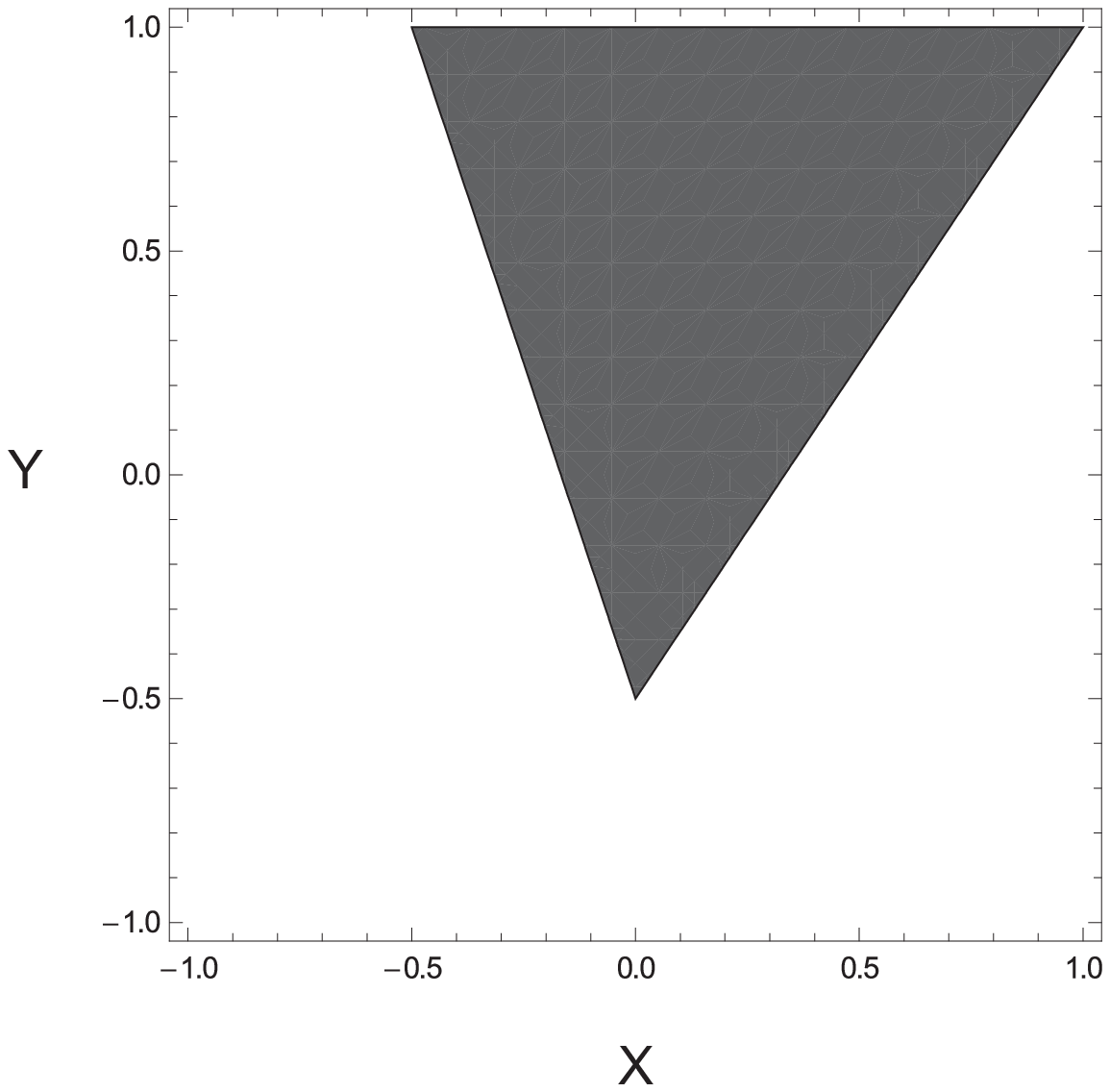}
\caption{The dark region in the figure shows these values of
parameters $\Lambda_3=\Lambda_8=Y,\;\Lambda_{i\neq 3,8}=X$ that
satisfy CP conditions. The qutrit channel has only the diagonal
(damping) part.} \label{two}
\end{figure}

\begin{figure}[h!]
\includegraphics[scale=0.4]{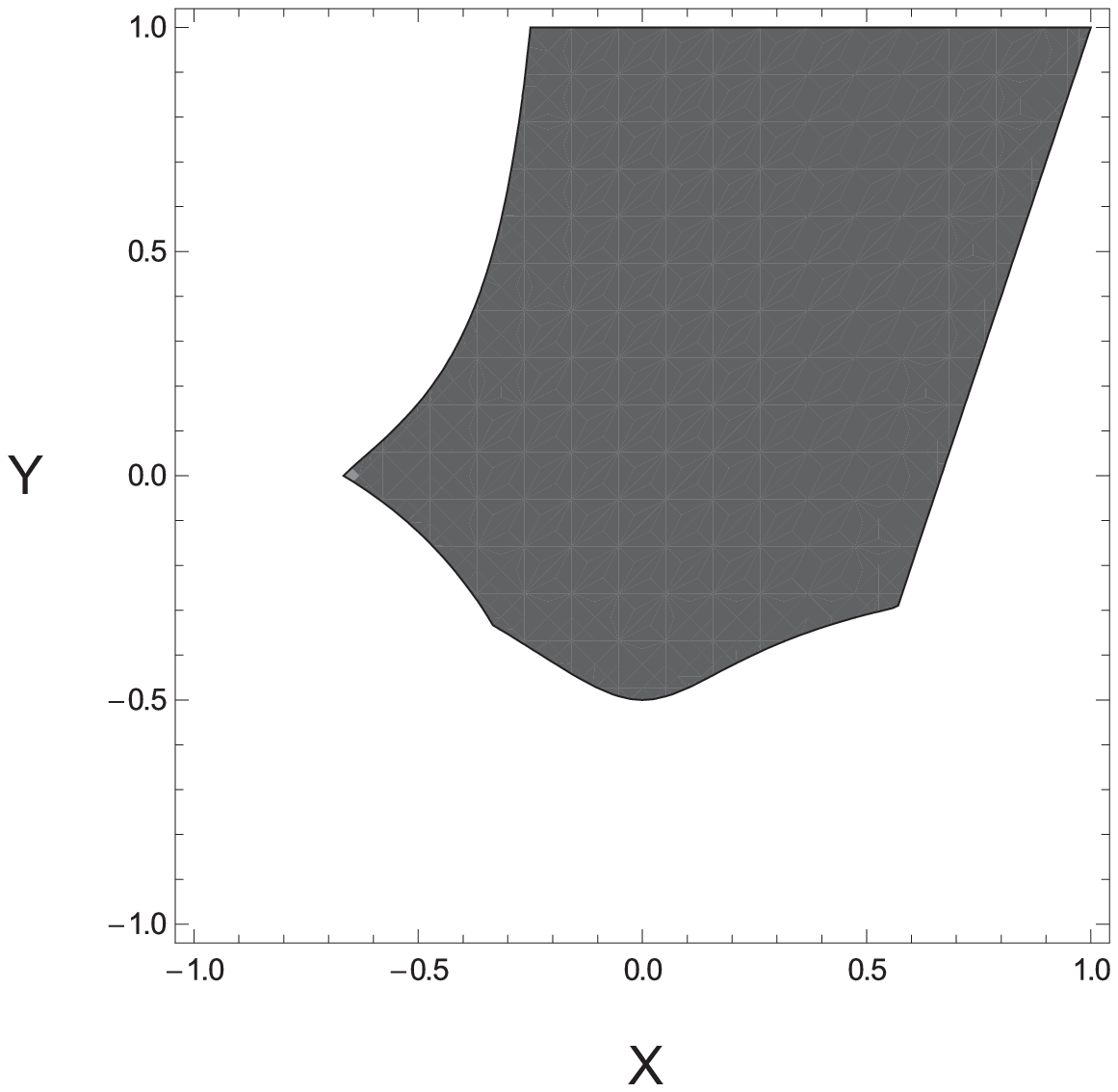}
\caption{The dark region in the figure shows these values of
parameters $\Lambda_3=X,\,\Lambda_8=Y,\;\Lambda_{i\neq 3,8}=XY$
that satisfy CP conditions. The qutrit channel has only the
diagonal (damping) part.} \label{three}
\end{figure}

\begin{figure}[h!]
\includegraphics[scale=0.4]{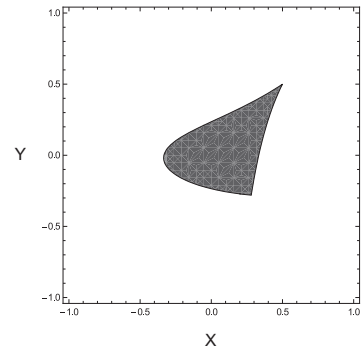}
\caption{The dark region in the figure shows these values of
parameters
$\Lambda_1=\Lambda_2=X,\;\Lambda_3=\Lambda_8=XY,\,\Lambda_{i\neq
1,2,3,8}=Y$ that satisfy CP conditions. The qutrit channel has
only the diagonal (damping) part.} \label{four}
\end{figure}

For qubit case, the allowed values of damping parameters
$\{\Lambda_{qb,i}^{CPM}\}_{i=1,2,3}$ form a characteristic
structure (tetrahedron, \cite{ruskai}). We are interested in the
structure that appears in qutrit case. The main obstacle here is
the size of parameter space. We have eight parameters on which we
impose our CPM constraints. We can investigate the
$\{\Lambda^{CPM}\}$ set projecting it onto subspaces. It is easier
to work with parameters paired according to $(\Lambda_1, \Lambda_2),\,
(\Lambda_4, \Lambda_5),\,(\Lambda_6, \Lambda_7)$ (this pairing
refers to the form of Gell-Mann matrices). We can also put
together $(\Lambda_3, \Lambda_8)$ (two diagonal Gell-Mann
matrices) though not necessarily, since the equations
are not symmetric in these two.\\
On Fig.\ref{one}-Fig.\ref{four} we show projections of
$\{\Lambda^{CPM}\}$ onto the various subspaces in 8 dimensional
space of parameters $\Lambda_1,...,\Lambda_8$. The \emph{dark
regions} in these figures correspond to these combinations of
$\Lambda_i$ which are \emph{satisfying CP conditions}. There are
many such projections that can be obtained from the conditions
that we have derived. In principle, the structure of the set of
$\{\Lambda^{CPM}\}$ is not simply a generalization of a
tetrahedron. Since we have three (or two) conditions for $\Lambda_i$ that
couple all the parameters (in a nonlinear way), the simple
polyhedron type structure (emerging from inequalities that are
linear in $\Lambda_i$, (\ref{cond1})) is altered. In the figures,
one can see combination of almost rough edges with smooth behavior
elsewhere.

\subsection{CPM conditions for channels based only on translations}

\begin{figure}[h!]
\includegraphics[scale=0.4]{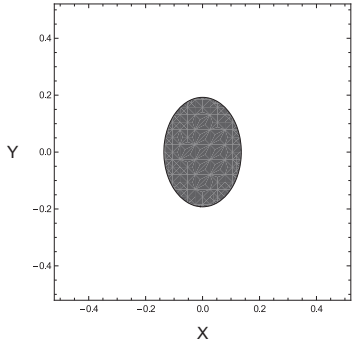}
\caption{The dark region shows these values of parameters
$X,\,Y$ that satisfy CP conditions, when translation acting on
qutrit Bloch vector has the form $T_1=(X,X,Y,0,0,0,0,0)$.}
\label{five}
\end{figure}

In this section we will analyze shortly the constraints of complete positivity on the possible translations.
The change of qutrit Bloch vector in this case will be of the form:
\begin{equation}
\vec{n}\mapsto \vec{n}'= \vec{n}+\vec{t},\label{trans}
\end{equation}
where $\vec{t}=(t_1,...,t_8)$. This type of channel is nonunital.
Below we show some examples of channels, for which we choose just two free parameters.
First, let us look at the translation of the form
\begin{equation}
T_1=(X,X,Y,0,0,0,0,0).
\end{equation}
 It turns out that effectively, parameters $X,\, Y$ must satisfy
\begin{equation}
1-3\sqrt{3}\sqrt{2X^2+Y^2}\,\ge\,0,
\end{equation}
what graphically is represented on Fig.\ref{five} - the dark
region corresponding to CP-allowed parameter values has an
ellipsoid form.

\begin{figure}[h!]
\includegraphics[scale=0.4]{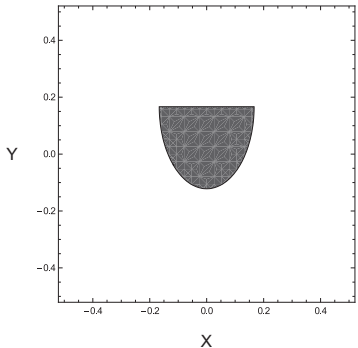} \caption{The dark region shows these values of
 parameters $X,\,Y$ that satisfy CP conditions, when translation acting on qutrit Bloch vector has the
form $T_2=(X,X,Y,0,0,0,0,Y)$.} \label{six}
\end{figure}

On the other hand, if we let the translation to shift also the 8th
component by the same amount as the 3rd component, therefore
translation having a form
\begin{equation}
 T_2=(X,X,Y,0,0,0,0,Y),
\end{equation}
then the allowed set of parameters $X,\,Y$ is further limited
with respect to $X$ - this parameter must satisfy $Y\leq1/6$.
And both parameters $X,\,Y$ must satisfy
\begin{equation}
1+3Y-3\sqrt{3}\sqrt{2X^2+Y^2}\,\ge\,0.
\end{equation}
This is shown on Fig.\ref{six}. The ellipsoid shape from
Fig.\ref{five} is now reshaped.

\subsection{CPM conditions for diagonal channels with translations}

\begin{figure}[h!]
\begin{center}
\includegraphics[scale=0.4]{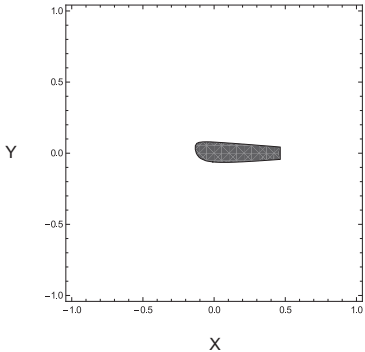}
\caption{The dark region in the figure shows these values of
parameters $\Lambda_i=X,\,t_i=Y$ that satisfy CP conditions. The
qutrit channel contains both the diagonal (damping) part and the
translation.}  \label{seven}
\end{center}
\end{figure}

We have seen what are the CP conditions for diagonal channels and
investigated some examples of channels built up only with
translations. What occurs when these two effects combine? Let us
take the channel that changes the Bloch vector according to
\begin{eqnarray}
\vec{n}&\mapsto& \vec{n}'= \Lambda\vec{n}+\vec{t},\nonumber\\
\rho&\mapsto&\Phi(\rho)=\frac{1}{3}(\mathbb{I}\,+\,\sqrt{3}(\Lambda\vec{n}+\vec{t})\cdot\vec{\lambda}).\label{all}
\end{eqnarray}
In this case, to evaluate positivity of $D_{\Phi}$ we evaluated
the principal minors of the matrix. In principle, the matrix is
positive when all the principal minors are positive, and it is
negative, when the principle minors have alternating signs.
However, it may happen that for some parameter values the
principal minors equal to zero and the method do not detect all
possible parametrization allowed by CP conditions. Nevertheless,
we use this method to analyze some cases and detect possible
regions of positivity of dynamical matrix $D_{\Phi}$. We do not
present here the list of inequalities corresponding to CP
conditions because their complexity would unable any insight into
the problem. We project the set of $\{\Lambda_i,t_i\}^{CPM}$ onto
some subspaces to gain a geometrical picture. Below, we can see
two examples. On Fig.\ref{seven} the dark region corresponds to
the CP-allowed values of $X,\,Y$, for which we assume that all
$\Lambda_i=X$ and $t_i=Y$.
\begin{figure}[h!]
\includegraphics[scale=0.4]{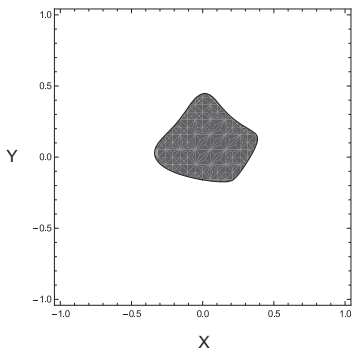}
\caption{The dark region in the figure shows these values of
parameters $\Lambda_{1,2}=X,\,\Lambda_{i\neq 1,2}=Y,\,t_{i}=XY$
that satisfy CP conditions. The qutrit channel contains both the
diagonal (damping) part and the translation.} \label{eight}
\end{figure}
Fig.\ref{eight} shows a different choice of parametrization and
relation between damping parameters and translation. We let
$\Lambda_{1,2}$ have independent value (X) from the rest of
$\Lambda_i$ (Y), and we assume that translation is equal to the
product of these two (XY).

\subsection{Two-qutrit states and affine transformations of qutrit Bloch vectors}

As already said, the dynamical matrix $D_{\Phi}$ corresponds to a
density matrix via $\rho_{\Phi}=\frac{1}{N}D_{\Phi}$. The latter,
in our case ($N=3$) describes a class of two-qutrit states that
can be parameterized by $\{\Lambda_i,\,t_i\}^{CPM}$. The two
qutrit state space is being investigated. Especially, the so
called magic simplex which can be considered an analog of the
magic tetrahedron of bipartite qubits \cite{baumgartner}. The
magic simplex of bipartite qutrits is only embedded in the space
of all bipartite qutrits. As an example, it does not contain a
state given by the density operator
\begin{equation}
\rho\,=\,\frac{1}{3}(|\Psi\rangle\langle\Psi|\,+\,2|\Phi\rangle\langle\Phi|),
\end{equation}
where $|\Psi\rangle\,=\,|0,0\rangle$ and
$|\phi\rangle\,=\,\frac{1}{\sqrt{2}}(|1,1\rangle\,+\,|2,2\rangle)$.
Interestingly, this state can be obtained from
$\rho_{\Phi}\,=\,\frac{1}{N}D_{\Phi}$ (N=3) with a proper choice
of parameters: $\Lambda_{3,6,7,8}\,=1$ and the rest equal to
0. A diagonal channel with such parameter values will transform
any qutrit Bloch vector according to
\begin{equation}
\vec{n}\,\longrightarrow\,\vec{n}'=\{0,0,n_3,0,0,n_6,n_7,n_8\}.
\end{equation}
Geometrically, the channel projects the Bloch vector onto the 3-6-7-8
subspace and the other components of $\vec{n}$ are lost. It resembles
therefore a phase flip type channel \cite{chuang}.

\section{Summary}

Basing on the generalized Bloch formalism for qutrit quantum
states we have analyzed quantum qutrit channels that have a form
of affine transformations on Bloch vectors. The aim was to derive
complete positivity conditions on channel parameters that may
appear in equations of evolution of the Bloch vector. We analyzed
diagonal channels (only with damping coefficients), for which we
obtained CP conditions on parameters in form of inequalities.
Analogous inequalities appear in qubit case - for which CP-allowed
channel parameters form a tetrahedron structure. The structure of
the corresponding set in qutrit case is more sophisticated, and
reveals not only polyhedron like characteristics. We analyzed also
channels which allow only shifting the Bloch vector, in which case
we investigated some specific examples of translations and the
corresponding CP constraints. The combined effect of damping and
shifting qutrit Bloch vector (diagonal channels with translation)
was also presented by projecting the CP-allowed set of channel
parameters onto specific subspaces. At the end we looked at the
two qutrit states that correspond to qutrit channels we
investigated, via Jamiolkowski isomorphism. As an example we give a 
channel that corresponds to a state which does not
belong to the so called magic simplex.\\
One of the interesting points would be to establish the relation
between the set of two qutrit states given by the dynamical matrix
that we analyze and the magic simplex for qutrits. Also the
analysis of the structure of CP-allowed channel parameters with
respect to entanglement breaking properties could reveal some
intriguing results.

\subsection*{Acknowledgements}
A. Checinska would like to thank R. Augusiak and M. Hall for
interesting comments.\\
This paper was supported by a MNiSW grant
No. 1P03B13730 and EMALI Marie-Curie Research Training Network. 

\appendix*
\section{}

In the discussion of the diagonal (damping) qutrit channels we
presented a set of conditions (\ref{cond1},\ref{cond2}) that come
from imposing positivity condition on $D_{\Phi}$. They correspond
to eigenvalues of the matrix, and three of them have sophisticated
form. In qutrit case, the dynamical matrix $D_{\Phi}$ is nine
dimensional. We already showed 6 eigenvalues. The remaining three
correspond to finding roots of polynomial of the 3rd order. Since
$D_{\Phi}$ is hermitian, all its eigenvalues must be real, to
guarantee complete positivity of the channel, they must also be
nonnegative. The polynomial $P(x) = Ax^3 + Bx^2 + Cx + D$ which we
analyze in order to obtain the rest of CP conditions (eigenvalues
$d_{7,8,9}$) has coefficients

\begin{tiny}\begin{eqnarray}
A&=&8,\qquad B=-24(1+\Lambda_3+\Lambda_8),\nonumber\\
C&=& 18((\Lambda_3+\Lambda_8)^2-(\Lambda_1+\Lambda_2)^2-(\Lambda_4+\Lambda_5)^2-(\Lambda_6+\Lambda_7)^2)+\nonumber\\
&\  &+24(1+2\Lambda_3+2\Lambda_8+\Lambda_3\Lambda_8),\nonumber\\
D&=&-8-18((\Lambda_3+\Lambda_8)^2-(\Lambda_1+\Lambda_2)^2-(\Lambda_4+\Lambda_5)^2+\nonumber\\
&\
 &-(\Lambda_6+\Lambda_7)^2)+27\Lambda_3((\Lambda_4+\Lambda_5)^2+(\Lambda_6+\Lambda_7)^2)+\nonumber\\
&\  &-54(\Lambda_1+\Lambda_2)(\Lambda_4+\Lambda_5)(\Lambda_6+\Lambda_7)+\nonumber\\
&\  &+9\Lambda_8(4(\Lambda_1+\Lambda_2)^2+(\Lambda_4+\Lambda_5)^2+(\Lambda_6+\Lambda_7)^2)+\nonumber\\
&\  &-24(\Lambda_8+\Lambda_3+\Lambda_3\Lambda_8)-4\Lambda_8(\Lambda_3+\Lambda_8)^2+\nonumber\\
&\  &-32\Lambda_8\Lambda_3^2-20\Lambda_8^2\Lambda_3.\nonumber
\end{eqnarray}
\end{tiny}

The roots can be of course found explicitly, but we do not present
them here. For some specific parameter values, number of real
roots can be reduced (and therefore number of CP inequalities). To
deduce that one has to analyze function
\[
f(a,b,c)=3^{-6}(3b-a^2)^3+54^{-2}(9ab-27c-2a^3)^2,
\]
and evaluate it at $f(B/8,C/8,D/8)$ (where $B,C,D$ are polynomial
coefficients given above). For the dynamical matrix $D_{\Phi}$
this function is always nonpositive (and indicates therefore real
roots). When $f(B/8,C/8,D/8)=0$ the polynomial has three real
roots and at least two are equal - then the number of CP
conditions reduces. This occurs for parametrization shown in
Fig.\ref{two}.

\end{document}